\documentclass[pageno]{jpaper}

\usepackage[normalem]{ulem}
\usepackage{subfig}[margin=10pt]
\usepackage{graphicx}
\usepackage{amsfonts}
\usepackage{amssymb}    
\usepackage{amsmath}
\usepackage{tablefootnote}
\usepackage{xurl}
\usepackage{balance}
\definecolor{a}{rgb}{0.83, 0.83, 0.83}
\definecolor{b}{rgb}{0.99,0.99,0.99}

\begin{document}
\author{James T. Meech \\ jtm45@cam.ac.uk \\ University of Cambridge}
\title{Is This Computing Accelerator Evaluation Full of Hot Air?}
\date{}
\maketitle

\thispagestyle{empty}

\begin{abstract}
Computing accelerators must significantly improve at least one metric such as: cost, speed, or efficiency for customers to find them useful. 
They must do this for at least one application that large numbers of users care about to become a commercial success. 
Furthermore, accelerators must improve the metric that customers care most about for a given application. 
We show that it is trivial to build an embedded sensor system that improves temperature sampling speed by $38,142\times$ and energy efficiency by $2\times$ compared to the state of the art in an application where these metrics do not matter. We then explain that this computing accelerator is not likely to displace the bimetallic kettle switch in the market because consumers are optimizing for cost, not temperature measurement sampling speed and efficiency when they buy a kettle.
\end{abstract}

\section{Introduction}
\label{section:introduction}
\begin{figure}[t]
    \centering
    \subfloat[\centering Image of commemorative stamp.]{{\includegraphics[height=4.9cm]{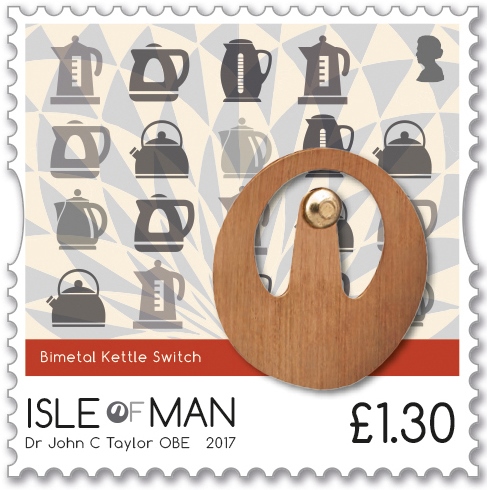}\label{fig:stamp}}}%
    \qquad
    \subfloat[\centering Moore state machine.]{{\includegraphics[height=4.9cm]{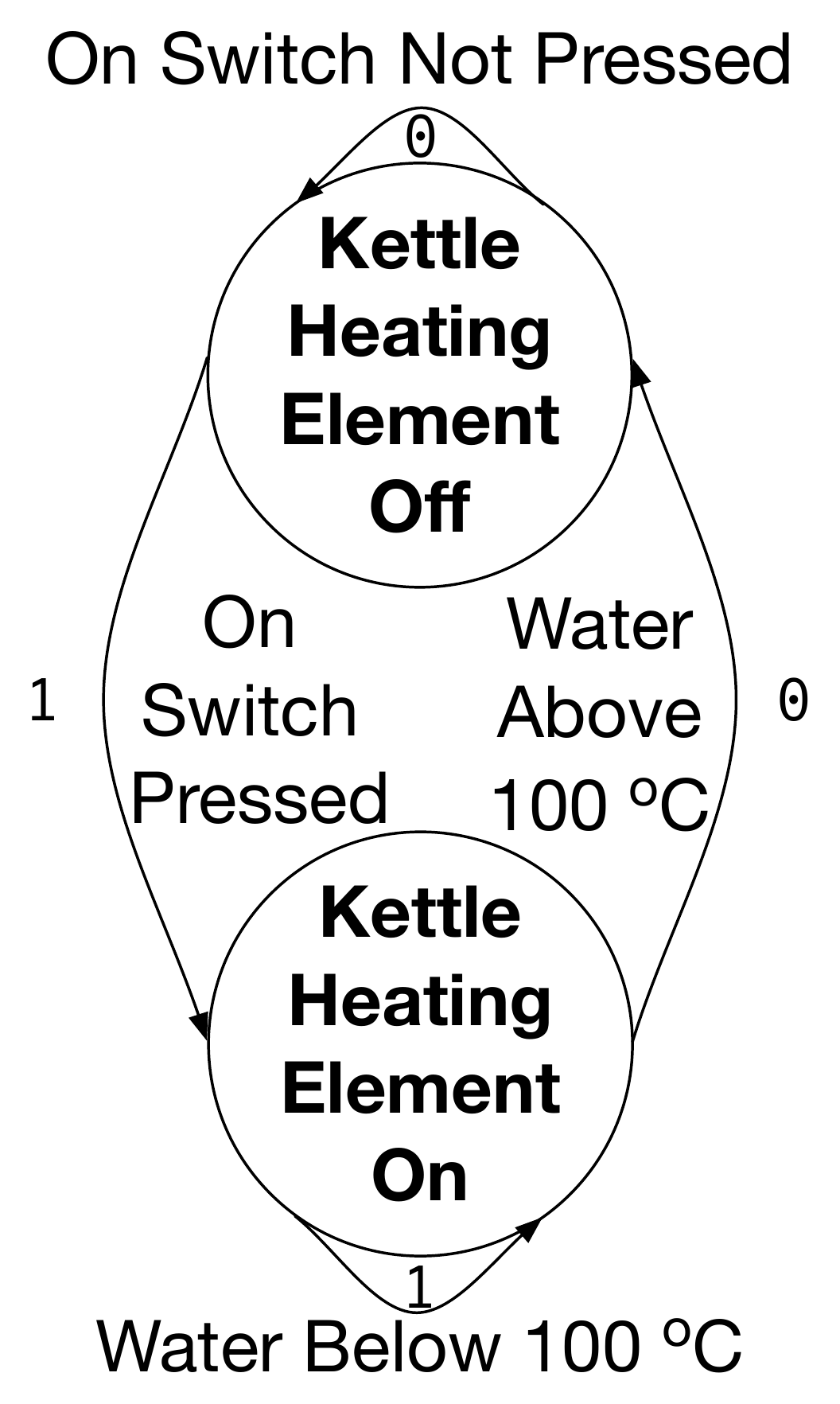}\label{fig:moore}}}%
    \caption{The stamp shows the bimetallic switch and the kettles that use it as a switch\cite{stamp}. The 1.30 GBP price is for the stamp but the bimetallic switch costs 0.83 GBP~\cite{thermostat}. The state machine of the kettle application switches on the kettle heating element when the kettle is switched on and continuously samples the water temperature until it reaches \unit{100}{\celsius}. 
	At this point it switches off the power supply to the kettle heating element.}%
    \label{fig:example}%
\end{figure}

We need new computing accelerator designs as they often accelerate research progress in the applications that run over them~\cite{hooker2021hardware}.
Researchers have built computing accelerators that achieved huge improvements over the state of the art by throwing out the traditional silicon transistor based digital computing stack. 
Examples include the Google quantum supremacy experiment ($1.6 \times 10^{9}$ $\times$ speedup), analog processing in memory ($2360 \times$ simulated performance improvement), and optical transformers ($8000 \times$ predicted energy efficiency improvement)~\cite{arute2019quantum, chi2016processing, anderson2023optical}.
It follows that other similar opportunities to use physics to perform analog sensing, computation, and actuation for free are waiting to be discovered by computer systems researchers. 
To illustrate this point we discuss the bimetallic kettle switch, an invention that performs a simple analog computing application without using a single silicon transistor.
Figure~\ref{fig:stamp} shows the bimetallic kettle switch that John C. Taylor invented in 1968~\cite{patent}.
A bimetallic switch is a transducer that can convert low-grade heat energy into mechanical switch movement. 
This makes it useful for mechanically switching off the electricity supply to a circuit once the switch reaches a threshold temperature. 
One of the most popular applications of the bimetallic switch is switching off the kettle (an electric appliance with the sole purpose of heating water) when the water inside it reaches boiling point (\unit{100}{\celsius}). 
John C. Taylor estimates that over two billion bimetallic switches have been sold since he invented them~\cite{sales}.
The bimetallic switch is a remarkable computing device, it makes continuous analog measurements of the temperature of the water in the kettle until it reaches \unit{100}{\celsius} and then immediately switches off a high voltage and high current power supply. It uses energy harvesting from the hot water to perform the switching action. 
Any traditional embedded sensor system designed to perform this temperature monitoring task would require a temperature sensor, microcontroller, power transistor, and an electromagnetic relay. The embedded system would therefore be significantly more costly than the bimetallic switch. 
Figure~\ref{fig:moore} shows a Moore state machine for the kettle application. In Section~\ref{section:comparison} we will benchmark an embedded sensor system that we designed to perform the function of this state machine against the bimetallic switch.

\section{Energy, Speed, and Cost Comparison}
\label{section:comparison}
We benchmarked an embedded sensor system designed to perform the same function as the bimetallic switch against analytical calculations of the maximum theoretical temperature sampling speed and energy efficiency of the bimetallic switch. 
For our analytical calculations we modeled the bimetallic switch as two \unit{20}{\milli \meter} diameter, \unit{0.5}{\milli \meter} thick disks of copper and stainless steel welded together. 
Let $m$ be the mass of the bimetallic switch (\kilo\gram), $c$ be the specific heat capacity of the materials that the switch is made of (\joule \per \kilo\gram \kelvin), and $\Delta T$ the temperature change (\unit{80}{\celsius}) that we have to subject the bimetallic switch to, causing it to switch. Then $Q$ (\joule) is the heat energy that the water in the kettle has to supply to the bimetallic switch to cause it to switch where

\begin{equation}
	Q = m c \Delta T.
	\label{equation:heat}
\end{equation}

We use copper with specific heat capacity \unit{385}{\joule \per \kilo \gram \kelvin}, thermal conductivity \unit{385}{\watt \per \meter \kelvin}, and density $8.93 \times 10^{3}$\,\kilo\gram\per\meter\cubed~\cite{copper}. We use steel with specific heat capacity \unit{500}{\joule \per \kilo \gram \kelvin}, thermal conductivity \unit{16.2}{\watt \per \meter \kelvin}, and density $8.00 \times 10^{3}$\,\kilo\gram\per\meter\cubed~\cite{steel}.
Let $R_\mathrm{T}$ be the thermal resistance (\kelvin \per \watt) and $C_\mathrm{T}$ be the thermal capacitance (\joule \per \kelvin) of the bimetallic switch.
We can calculate the sampling speed that the bimetallic switch can operate at by calculating its thermal time constant $\tau$ (\second) where

\begin{equation}
	\tau = R_\mathrm{T} C_\mathrm{T}
	\label{equation:resistance}
\end{equation}

and

\begin{equation}
	C_\mathrm{T} = mc.
	\label{equation:capacitance}
\end{equation}

Using the bimetallic switch dimensions and material properties stated in this section, the energy required for one run of the kettle switching application is \unit{90.79}{\joule}.
Table~\ref{table:costs} shows that the embedded sensor system that we designed cost 31.45 GBP in one off quantity. Therefore, the bimetallic switch provides a $38\times$ cost reduction compared to the embedded sensor system. 
MacKay states that a kettle consumes \unit{3.00}{\kilo \watt} of power when switched on~\cite{mackay2008sustainable}.
We measured that a kettle takes approximately \unit{210}{\second} to boil and therefore our embedded sensor system would consume \unit{45.52}{\joule} of energy sampling the temperature of the water whilst waiting for the kettle to boil. 
Using Equation~\ref{equation:heat} we estimated that the bimetallic switch consumes $2\times$ more energy than our prototype embedded system but we do not care because the embedded sensor system consumes 0.0072\,\% of the power that we are dissipating in the kettle heating element~\cite{Arnaud2015}.
Therefore, our embedded sensor system is $2 \times$ more efficient than the bimetallic switch.  
We measured that our Microchip ATmega328 based embedded temperature sensor system uses \unit{0.217}{\watt} of power using a Microchip PAC1934 USB power meter. 
We measured the sensor sample rate to be \unit{8928.57}{\hertz} by measuring the time it took to sample the temperature 1,000,000 times using a millisecond timer. 
Using Equations~\ref{equation:resistance}~and~\ref{equation:capacitance} we calculated that the bimetallic switch has a thermal time constant of \unit{4.27}{\second} and therefore a maximum sample rate of \unit{0.234}{\hertz} which is $38,142\times$ slower than our embedded sensor system. 

Our embedded accelerator improves over the state-of-the-art bimetallic switch on two out of the three metrics we evaluate it against. 
It is more than four orders of magnitude better on sample rate. Does this mean that we should immediately patent the embedded sensor system and start a company that sells kettles which use our embedded sensor system computing accelerator?
The answer is no. The users of the kettle application care much more about cost than they do about energy efficiency or sensor sampling rate. 

\section{Accelerate the Whole Application}

The bimetallic switch is able to achieve such great cost improvements over embedded sensor system because it runs the entire kettle application. 
If it was only able to run part of the application it would need a digital electronic coprocessor to run the rest of the application, therefore, increasing the cost of the system. 
From Ahmdal's law~\cite{amdahl1967validity} we know that any accelerator which accelerates a fraction $A$ of a program can only produce a maximum theoretical speedup of $\frac{1}{A}$.

\begin{table}[t]
\caption{The cost of each component in the embedded sensor system designed to replace the bimetallic switch in a kettle.}
\resizebox{\columnwidth}{!}{%
\begin{tabular}{ccc}
\toprule
\rowcolor{a} Component                                                                                        & Cost (GBP) & Source           \\ 
\midrule
\rowcolor{b} Atmega328 Evaluation Board\tablefootnote{Yes, the Atmega328 evaluation board is an Arduino Uno.} & 23.38      & \cite{Atmega328} \\ 
\rowcolor{a} LM35DZ Temperature Sensor                                                                        & 1.15       & \cite{LM35DZ}    \\ 
\rowcolor{b} 2N3904 NPN Transistor                                                                            & 0.07       & \cite{2N3904}    \\ 
\rowcolor{a} PT270730 Electromechanical Relay                                                                 & 6.85       & \cite{PT270730}  \\ 
\bottomrule
\end{tabular}}
\label{table:costs}
\end{table}

\section{Improve Metrics That People Care About}

The bimetallic switch is a commercial success because it optimizes for cost in an application that people care about. 
Many people around the world make multiple hot beverages each day and require low-cost equipment to help them do this fast and efficiently. 
The computer systems community should constantly search for impactful applications for application-specific computing accelerators. 
We can sometimes unlock orders of magnitude improvements in the metrics that matter by throwing out silicon transistors and the entire digital computing stack that we have built around them.
Computer systems researchers are well placed to realize these improvements as they occupy a higher abstraction level than emerging computing device researchers~\cite{tye2021bridging}.
Harnessing the inherent randomness in new or emerging computing devices to build new computer architectures that are faster and more efficient for running Monte Carlo simulations is a promising example of such an application. Zhang et al. and the COINFLIPS project team have hinted that we can harness the physical properties of these novel devices to accelerate entire applications~\cite{zhang2021statistical,shashank2022magnetic}.
Unfortunately many researchers, including the author have fallen into the trap of building computing accelerators that do not accelerate the important metric or that do not produce significant improvements over the state of the art.
We conclude with a checklist for a good computing accelerator to help the research community avoid building accelerators that nobody cares about.
Good accelerators will tick all boxes.

\section*{Checklist for a Good Computing Accelerator}

\begin{enumerate}[itemsep=2pt]
    \item[$\square$] \begin{minipage}[c]{\columnwidth}{Does the accelerator significantly improve upon a important metric such as speed, efficiency, or cost?}\end{minipage} 
    \item[$\square$] \begin{minipage}[c]{\columnwidth}{Does the accelerator produce the significant improvement for an application that users care about?}\end{minipage} 
    \item[$\square$] \begin{minipage}[c]{\columnwidth}{Is the metric that the accelerator improves upon a metric that users care about for this application?}\end{minipage} 
\end{enumerate}

\balance

\bibliographystyle{plain}
\bibliography{references}

\end{document}